# Isospin Decomposition of D Mesons

Shakeel Mahmood

*Department of Physics, Air University,*
*PAF Complex, E-9, Islamabad, Pakistan.*
*E-mail:shakeel.mahmood@mail.au.edu.pk*

Mudasir Hussain

*Department of Physics, Air University,*
*PAF Complex, E-9, Islamabad, Pakistan.*
*E-mail:191619@students.au.edu.pk*

Abrar Ahmad

*Department of Physics, COMSATS University Islamabad, Pakistan.*
*Islamabad, Pakistan. E-mail:smart5733@gmail.com*

This work focuses on decomposition of isospin amplitude of D meson non-leptonic decays. Isospin vector algebra is used to show the equivalency of two amplitude decompositions. We restrict to the transitions involving only $\Delta I = 1$ and $\Delta I = 0$. The isospin symmetry is relating charge channel ($D^+ \rightarrow K^+\pi^+\pi^-$, $D^+ \rightarrow K^+\pi^0\pi^0$ and $D^+ \rightarrow K^0\pi^0\pi^+$), and neutral channel ($D^0 \rightarrow K^0\pi^+\pi$, $D^0 \rightarrow K^0\pi^0\pi^0$ and $D^0 \rightarrow K^+\pi^0\pi^-$) with each other. Equivalent triangle relations are obtained for both channels.

## 1. INTRODUCTION

The Cabibbo-Kobayashi-Meskawa (CKM) matrix [1, 2] describe the mixing of quarks and CP violation. Many B and D decays are measuring phases of CKM matrix [3]. CP violation in K and B meson is large as compared to D meson. We need very clean channels for the search of CP violation in D meson. This is the purpose of this article to identify clean channels free of any strong CP contamination.

The isospin amplitudes decomposition for $B \rightarrow \overline{D}^{(*)}D^{(*)}K$ decays have been utilized for the extraction of strong CP phase with experimental data available [5, 6]. Similar treatment is also possible for D meson decays which is ignored sector for CP violation as compared to B meson. The elimination of hadronic uncertainties coming from penguin diagrams pointed by [7] and [8] . They showed that these uncertainties can be eliminated by the using of isospin decomposition, if certain set of reactions are making a closed figure. The CP violation in D meson is very small as compared to B decays, so any minor effect should not be ignored. In the present article we propose isospin amplitude decomposition of cabibo favoured $D \rightarrow K\pi\pi$ reaction and use it for isospin amplitude decomposition. Here the contribution from flavour changing neutral current have two loops, so obviously making it small as compared to tree level contribution. But this small effect can pollute already small CP violation in D mesons. So exploration of $D \rightarrow K\pi\pi$ this purpose is important. Direct CP violation through interference between tree level and loop level diagram was pointed by [9] for $D \rightarrow \pi\pi\pi$. In the case of $D \rightarrow \pi\pi\pi$ both tree and loop level contribution are on equal footing but for our reactions $D \rightarrow K\pi\pi$, interference is very small. So no direct CP violation can be searched but large branching ratio [10] is making it attractive experimentally. As pointed out by FOCUS Colleberation [11] this three body decay is contaminated with two body $D \rightarrow K\pi$ decay and a separate isospin phase. these problems can be tackled by using isospin amplitude decomposition.

After this brief introduction Hamiltonian of the reaction is discussed in section 2. In section 3 amplitude transition is given. Section 4 discusses the isospin algebra involved. Isospin decomposition is studied in section 5. Triangular relations are established in section 6 and finally summary is provided in section 7.

## 2. HAMILTONIAN FORMULATION

The weak hamiltonian reponsible for $D \rightarrow K\pi\pi$ decay can be written as

$$\hat{H}_W = \frac{G_F}{\sqrt{2}} V_{us} V_{cd} (\overline{u}\gamma_\mu(1-\gamma_5)s)(\overline{d}\gamma_\mu(1-\gamma_5)c) + h.c. \tag{1}$$

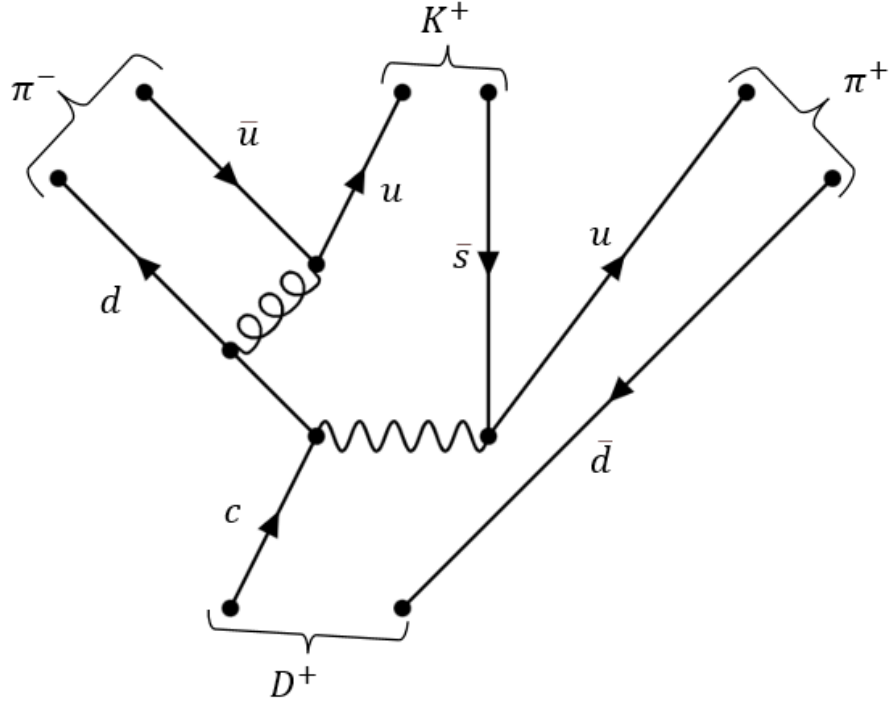

FIG. 1: Tree Level Feynman Diagram of $D^+ \to K^+ \pi^+ \pi^-$

Isospin ket of first term involved in Hamiltonian Eq. 1[$\overline{u}s$] and [$\overline{d}c$] are

$$\left|\frac{1}{2}, -\frac{1}{2}\right\rangle$$

and

$$\left|\frac{1}{2}, +\frac{1}{2}\right\rangle$$

respectively. So combining isospin of both operators,we can construct isospin triplet $|1,0\rangle$and isospin singlet $|0,0\rangle$ given in Eq. 2

$$\left|\frac{1}{2}, +\frac{1}{2}\right\rangle \left|\frac{1}{2}, -\frac{1}{2}\right\rangle = \frac{1}{\sqrt{3}}\,|1,0\rangle + \frac{1}{\sqrt{2}}\,|0,0\rangle\,. \tag{2}$$

So Hamiltonian can be written as a sum of two Hamiltonians

$$\hat{H}_W = \hat{H}_0 + \hat{H}_I$$

and in Dirac notation $\left|\hat{H}_0\right\rangle = |0,0\rangle$ and $\left|\hat{H}_I\right\rangle = |1,0\rangle$.

Feynman diagrams of $D^+ \to K^+ \pi^- \pi^+$are shown in Fig. 1 and Fig. 2

## 3. AMPLITUDE OF TRANSITION

The isospin amplitude of transition of $D \to K\pi\pi$ with above Hamiltonian is given in Eq. 3

$$\begin{aligned}
A(D \;\to\; K\pi\pi) &= \langle K\pi\pi|\,\hat{H}_0\,|D\rangle + \langle K\pi\pi|\,\hat{H}_I\,|D\rangle \\
&= \langle I_K + I_\pi + I_\pi\,|I_D + I_{H_0}\rangle + \langle I_K + I_\pi + I_\pi\,|I_D + I_{H_I}\rangle\,.
\end{aligned} \tag{3}$$

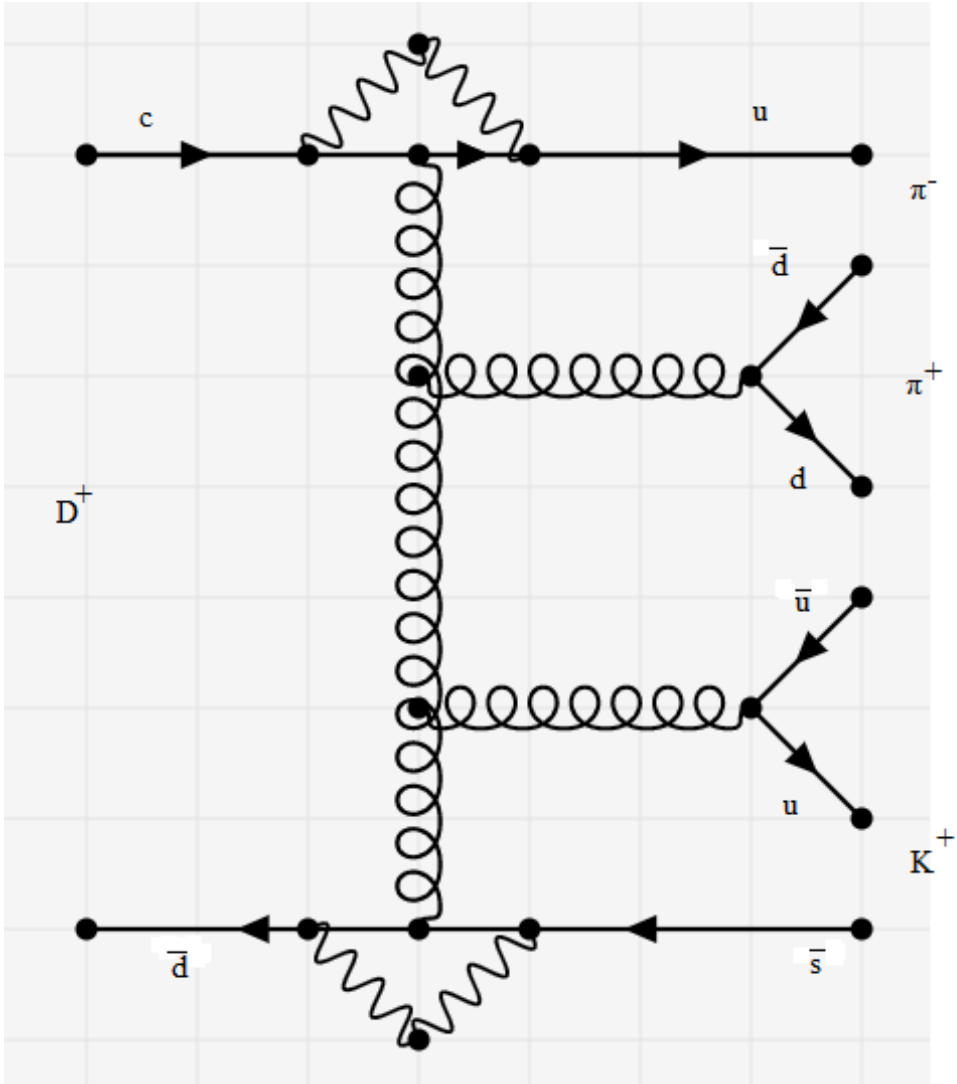

FIG. 2: Penguin Diagram of $D^+ \rightarrow K^+ \pi^+ \pi^-$

Here $|I_D + I_{H_0}\rangle$ and $|I_D + I_{H_I}\rangle$ are obtained by vectorially adding isospin of $D$ meson and isospin of $H_0$ and $H_I$ respectively. Similarly, $\langle I_K + I_\pi + I_\pi|$ obtained by vectorial addition of isospin of $K$ and $\pi$ mesons.

## 4. ISOSPIN ADDITION

Addition of isospin is done by the use of formalism of angular momenta addition used in quantum mechanics [4]. As two isospin states are involved in the initial state, so total isospin is vectorial sum of $I_1$ and $I_2$, i.e., Eq. 4

$$I = I_1 + I_2. \tag{4}$$

The commutation relations,

$$[I, I_z] = 0, [I_1, I_{1z}] = 0, [I_2, I_{2z}] = 0 \text{ but } [I, I_{1z}] \neq 0 \text{ and } [I, I_{2z}] \neq 0$$

are leaving us only two choices of complete set of commuting operators;

I: $I_1^2, I_2^2, I_{1z}, , I_{2z}$ and

II: $I_1^2, I_2^2, , I, , I_z$.

In terms of eigen values of angular momentum operators we can write uncoupled representation $|j_1 j_2 m_1 m_2\rangle = |j_1 m_1\rangle |j_2 m_2\rangle$ and coupled representation $|j_1 j_2 jm\rangle$. These two representations are linked with each other by the use of Clebsch-Gordon coefficients as given below Eq. 5

$$|j_1 j_2 jm\rangle = \sum_{m_1 m_2} \begin{bmatrix} j_1 & j_2 & j \\ m_1 & m_2 & m \end{bmatrix} |j_1 j_2 m_1 m_2\rangle. \tag{5}$$

For three isospins $I = I_1 + I_2 + I_3$,the uncoupled representation state is shown in Eq. 6

$$|j_1 j_2 j_3 m_1 m_2 m_3\rangle = |j_1 m_1\rangle |j_2 m_2\rangle |j_3 m_3\rangle \tag{6}$$

generated by the set of commuting operators $(I_1^2, I_2^2, I_3^2, I_{1z}, ., I_{2z}, I_{1z})$.

For the coupled representation, we have to add any two first and then add third one, i.e., $I_{12} = I_1 + I_2$ and $I = I_{12} + I_3$ with commuting operators $I_1^2, I_2^2, I_3^2, I_{1z}, ., I_{2z}$, and other case is $I_{23} = I_2 + I_3$ and $I = I_1 + I_{23}$. Corresponding to these situations we have two different sets of commutation operators (CSCO) $I_1^2, I_2^2, I_3^2, I_{12}^2, I^2, I_z$ and $I_1^2, I_2^2, I_3^2, I_{23}^2, I^2, I_z$. These two sets of CSCO have different coupled representations of bases states $|j_{12} j_3; jm\rangle$ and $|j_1 j_{23}; jm\rangle$ respectively. These are linked with uncoupled bases states as in 7 and 8

$$|j_{12} j_3; jm\rangle = \sum_{m_1' m_2'} \begin{pmatrix} j_1 & j_2 & j_{12} \\ m_1 & m_2 & m_{12} \end{pmatrix} \begin{pmatrix} j_{12} & j_3 & j \\ m_{12} & m_3 & m \end{pmatrix} |j_1 m_1\rangle |j_2 m_2\rangle |j_3 m_3\rangle \tag{7}$$

$$|j_1 j_{23}; jm\rangle = \sum_{m_1 m_2} \begin{pmatrix} j_2 & j_3 & j_{23} \\ m_2' & m_3' & m_{23} \end{pmatrix} \begin{pmatrix} j_1 & j_{23} & j \\ m_1 & m_{23} & m \end{pmatrix} |j_1 m_1'\rangle |j_2 m_2'\rangle |j_3 m_3'\rangle \tag{8}$$

The coupled representations are related to each other through unitary transformation.

## 5. ISOSPIN AMPLITUDE DECOMPOSITION

The ket $|I_D + I_H\rangle$ in the uncoupled representation is written as in Eq. 9

$$\begin{aligned} H_I |D^+\rangle &= \sqrt{\frac{2}{3}} \left| \frac{3}{2}, \frac{1}{2} \right\rangle - \sqrt{\frac{1}{3}} \left| \frac{1}{2}, \frac{1}{2} \right\rangle \\ H_0 |D^+\rangle &= \left| \frac{1}{2}, \frac{1}{2} \right\rangle \\ H_I |D^0\rangle &= \sqrt{\frac{2}{3}} \left| \frac{3}{2}, -\frac{1}{2} \right\rangle - \sqrt{\frac{1}{3}} \left| \frac{1}{2}, -\frac{1}{2} \right\rangle \\ H_0 |D^0\rangle &= \left| \frac{1}{2}, -\frac{1}{2} \right\rangle \end{aligned} \tag{9}$$

The state $|(I_K + I_\pi) + I_\pi\rangle$ is written in coupled bases $|j_{12} j_3; jm\rangle$ which are linked with uncoupled base Eq. 6. We obtain the bracket $\langle D| H |K\pi\pi\rangle_{I_{12}}$ whose amplitudes are given below in Eqs. 10,11 and 12 in terms of six amplitudes $\bar{a}_{\frac{3}{2}(\frac{1}{2})}$, $a_{\frac{3}{2}(\frac{1}{2})}'$ and $a_{\frac{3}{2}(\frac{1}{2})}$

$$A(D^+ \to K^0 \pi^+ \pi^0) = -\sqrt{\frac{1}{15}} \bar{a}_{\frac{3}{2}} - \sqrt{\frac{1}{3}} \bar{a}_{\frac{1}{2}} \tag{10}$$

$$\begin{aligned} A(D^+ \to K^+ \pi^0 \pi^0) = &-\frac{2}{3}\sqrt{\frac{1}{15}} \bar{a}_{\frac{3}{2}} + \frac{2}{3}\sqrt{\frac{1}{3}} \bar{a}_{\frac{1}{2}} \\ &+ \frac{1}{3}\sqrt{\frac{2}{3}} a_{\frac{3}{2}}' - \frac{1}{3}\sqrt{\frac{1}{3}} a_{\frac{1}{2}}' \\ &+ \frac{\sqrt{2}}{3} a_{\frac{3}{2}} - \frac{1}{3} a_{\frac{1}{2}} \end{aligned} \tag{11}$$

$$A(D^+ \rightarrow K^+\pi^+\pi^-) = -\frac{1}{3}\sqrt{\frac{2}{15}}\overline{a}_{\frac{3}{2}} + \frac{1}{3}\sqrt{\frac{2}{3}}\overline{a}_{\frac{1}{2}}$$
$$-\frac{2}{3}\sqrt{\frac{1}{3}}a'_{\frac{3}{2}} - \frac{1}{3}\sqrt{\frac{2}{3}}a'_{\frac{1}{2}}$$
$$+\frac{2}{3}a_{\frac{3}{2}} - \frac{\sqrt{2}}{3}a_{\frac{1}{2}} \tag{12}$$

The amplitudes for the other bracket $\langle D| \, H \, |K\pi\pi\rangle_{I_{23}}$ corresponding to the ket $|I_K + (I_\pi + I_\pi)\rangle$ are given in Eqs. 13, 14 and 15 , and expressed in terms of three amplitudes $\overline{a}_2$, $a'_0$ and $a_0$.

$$A(D^+ \rightarrow K^+\pi^+\pi^-) = -\frac{2}{3}\sqrt{\frac{1}{5}}\overline{a}_2 - \frac{\sqrt{2}}{3}a'_0 + \sqrt{\frac{2}{3}}a_0 \tag{13}$$

$$A(D^+ \rightarrow K^+\pi^0\pi^0) = -\frac{2}{3}\sqrt{\frac{21}{5}}\overline{a}_2 + \frac{1}{3}a'_0 - \frac{1}{\sqrt{3}}a_0 \tag{14}$$

$$A(D^+ \rightarrow K^0\pi^0\pi^+) = - - \sqrt{\frac{1}{5}}\overline{a}_2 \tag{15}$$

Similarly we can obtain same amplitudes for neutral decay mode decay modes $D^0 \rightarrow K^0\pi^+\pi^-$, $D^0 \rightarrow K^0\pi^0\pi^0$ and $D^0 \rightarrow K^{+-}\pi^0\pi^-$. Here we have relation between $\overline{a}_2$, $a'_0$, $a_0$ and $\overline{a}_{\frac{3}{2}(\frac{1}{2})}$, $a'_{\frac{3}{2}(\frac{1}{2})}$, $a_{\frac{3}{2}(\frac{1}{2})}$ as below in Eqs. 16, 17 and 18.

$$a_0 = \sqrt{\frac{2}{3}}a_{\frac{3}{2}} - \frac{1}{\sqrt{3}}a_{\frac{1}{2}} \tag{16}$$

$$a'_0 = \sqrt{\frac{2}{3}}a'_{\frac{3}{2}} - \frac{1}{\sqrt{3}}a'_{\frac{1}{2}} \tag{17}$$

$$\overline{a}_0 = \sqrt{\frac{2}{3}}\overline{a}_{\frac{3}{2}} - \frac{1}{\sqrt{3}}\overline{a}_{\frac{1}{2}} \tag{18}$$

## 6.  TRIANGLE RELATIONS

The isospin decomposition amplitudes for neutral as well as charged mesons are forming triangles like

$$\frac{1}{\sqrt{2}}A(D^+ \rightarrow K^+\pi^+\pi^-) + A(D^+ \rightarrow K^+\pi^0\pi^0) + A(D^+ \rightarrow K^0\pi^+\pi^0) = 0 \tag{19}$$

for $\langle D| \, H \, |K\pi\pi\rangle_{I_{12}}$ shown in Eq. 19 and similarly for the other bracket $\langle D| \, H \, |K\pi\pi\rangle_{I_{23}}$

$$\frac{1}{\sqrt{2}}A(D^+ \rightarrow K^+\pi^+\pi^-) + A(D^+ \rightarrow K^+\pi^0\pi^0) + A(D^+ \rightarrow K^0\pi^+\pi^0) = 0 \tag{20}$$

shown in Eq. 20 and Figure 3. Similar results exist for neutral channels $D^0 \rightarrow K^0\pi^+\pi^-$, $D^0 \rightarrow K^0\pi^0\pi^0$ and $D^0 \rightarrow K^{+-}\pi^0\pi$.

$$\frac{1}{\sqrt{2}}A(D^0 \rightarrow K^0\pi^+\pi^-) + A(D^0 \rightarrow K^0\pi^0\pi^0) + A(D^0 \rightarrow K^+\pi^0\pi^-) = 0$$

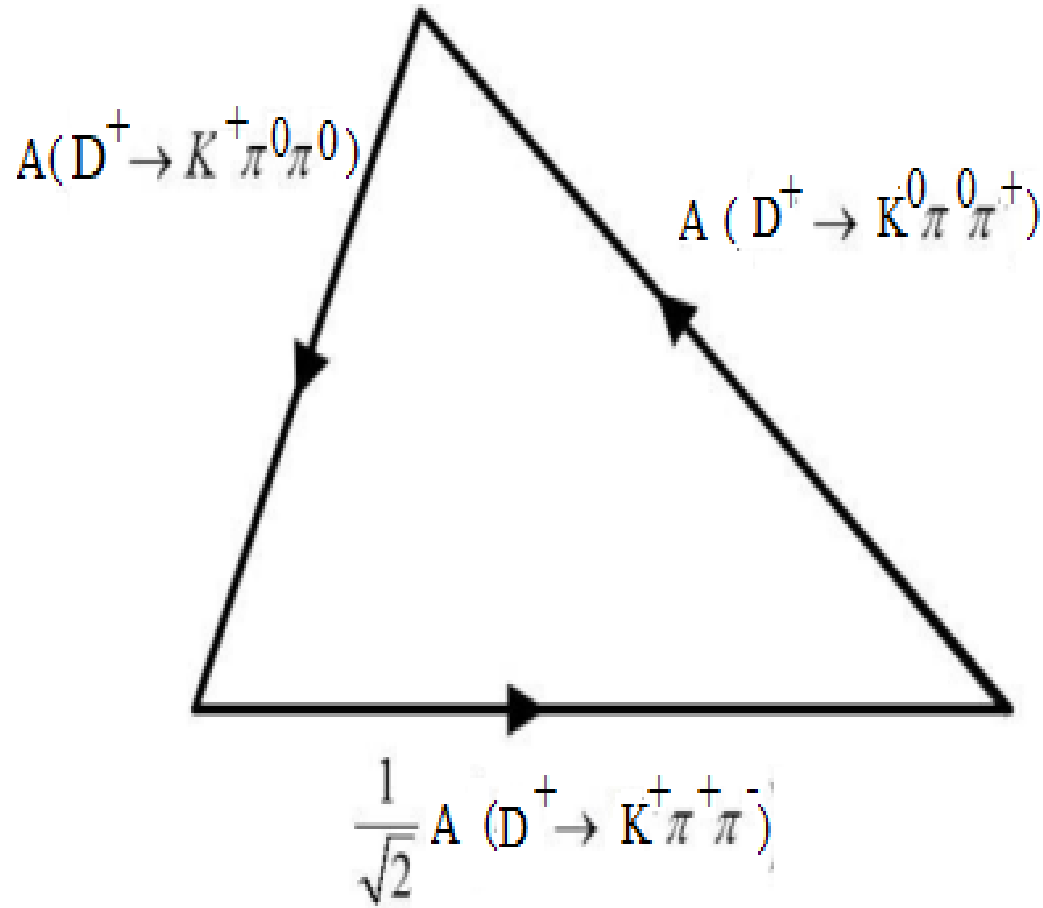

FIG. 3: Triangular relations $D \to K\pi\pi$ decays

## 7. SUMMARY

We have explored the three body non-leptonic D meson decay for isospin decomposition and obtained charge and neutral channel. Using quantum mechanics angular momentum addition we decompose isospin amplitude. These amplitudes exhibit triangular relationships for both neutral $D^0 \to K^0\pi^+\pi^-$, $D^0 \to K^0\pi^0\pi^0$ and $D^0 \to K^{+-}\pi^0\pi$ and charge channel $D^+ \to K^+\pi^+\pi^-$, $D^+ \to K^+\pi^0\pi^0$ and $D^+ \to K^0\pi^0\pi^+$. So these sets will be free of hadronic uncertainties when they are used for the study of CP violation.

---

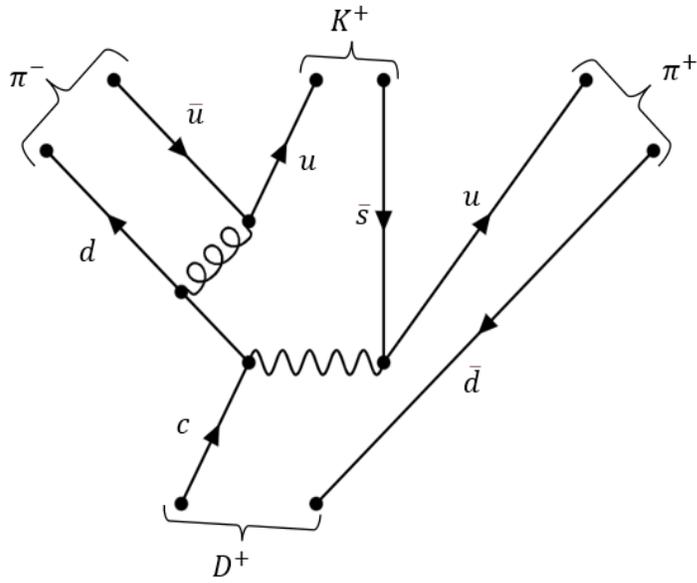

*Figure 1*

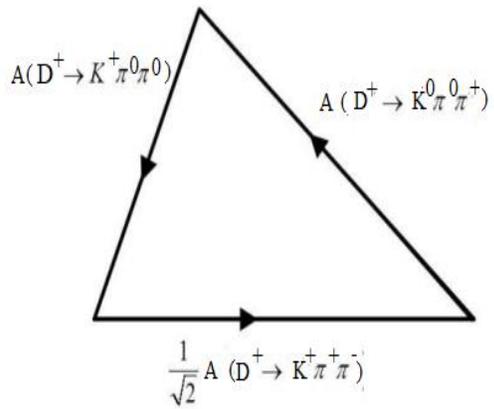

*Figure 2*

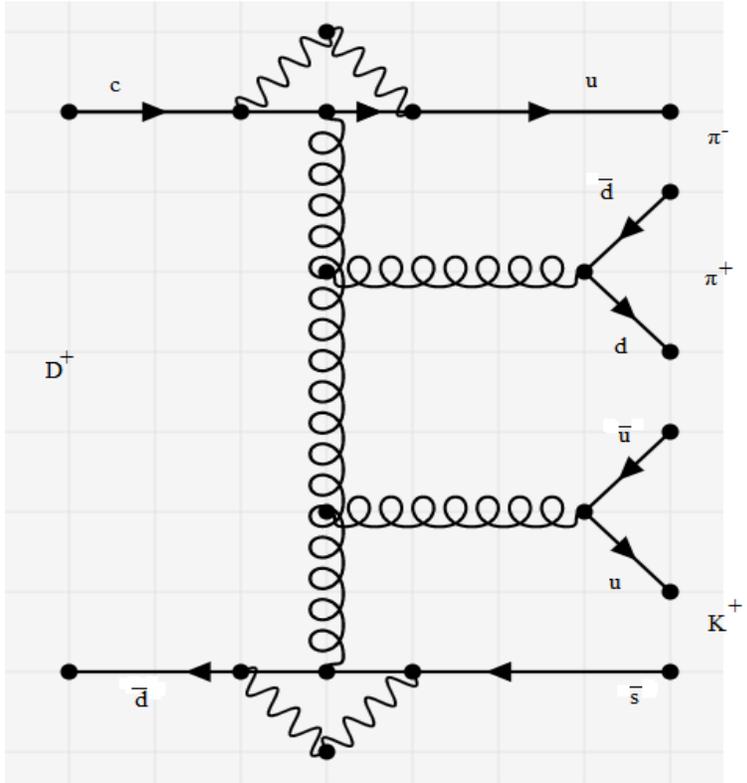

*Figure 3*